\let\latexaddtocontents\addtocontents

\documentclass[
 aip,
 cha,amsmath,amssymb,
preprint,
,titlepage]{revtex4-2}

\let\addtocontents\latexaddtocontents

\let\citeyear\relax
\let\textcite\relax

\let\citeauthor\relax

\expandafter\let\csname ver@natbib.sty\endcsname\relax

\usepackage{tabularx}
\usepackage{graphicx}
\usepackage{dcolumn}
\usepackage{dsfont}
\usepackage{bm}

\usepackage[utf8]{inputenc}
\usepackage[T1]{fontenc}
\usepackage{mathptmx}
\usepackage{etoolbox}

\usepackage{csquotes}
\usepackage[english]{babel}
\usepackage[backend=biber, 
    doi=true,style=phys]{biblatex}
\usepackage{xspace}
\usepackage{array}
\usepackage{stmaryrd}
\newcolumntype{K}[1]{>{\centering\let\newline\\\arraybackslash\hspace{0pt}}m{#1}}
\usepackage{makecell}
\usepackage{soul}

\makeatletter
\def\@email#1#2{%
 \endgroup
 \patchcmd{\titleblock@produce}
  {\frontmatter@RRAPformat}
  {\frontmatter@RRAPformat{\produce@RRAP{*#1\href{mailto:#2}{#2}}}\frontmatter@RRAPformat}
  {}{}
}%

\usepackage[mathscr]{euscript}
\usepackage{blkarray}
\usepackage{color}
\usepackage{xcolor}
\usepackage{xspace}
\usepackage{amsmath}
\usepackage{relsize}
\usepackage[bb=boondox]{mathalfa}
\usepackage{mathrsfs}
\usepackage{multirow}
\usepackage{mleftright}
\usepackage{mathtools}
\usepackage[super]{nth}
\DeclareMathOperator{\Tr}{Tr}

\usepackage{hyperref}
\hypersetup{
    colorlinks=true,
    citecolor=black,linkcolor=blue,
    filecolor=magenta,      
    urlcolor=cyan
    }

\makeatother

\begin{document}
\renewcommand{\bibliography}[1]{}

\title{
A quantum Pascal pyramid and an extended de Moivre-Laplace theorem}
\author{Mohamed Sabba\\ \href{mailto:m.sabba@soton.ac.uk}{m.sabba@soton.ac.uk}}
\affiliation{School of Chemistry, University of Southampton, SO17 1BJ, UK}

\date{\today}

\begin{abstract}
Pascal's triangle is widely used as a pedagogical tool to explain the "first-order" multiplet patterns that arise in the spectra of $I_N S$ coupled spin-1/2 systems in magnetic resonance. Various other combinatorial structures, which may be well-known in the broader field of quantum dynamics, appear to have largely escaped the attention of the magnetic resonance community with a few exceptions, despite potential usefulness.\par

In this brief set of lecture notes, we describe a "quantum Pascal pyramid" (OEIS \href{https://oeis.org/A268533}{A268533}) as a generalization of Pascal's triangle, which is shown to directly map the relationship between multispin operators of arbitrary spin product rank $q$ ($\hat{Z}_N^q$) and population operators for states with magnetic quantum number $m$ ($\hat{S}_N^m$), and - as a consequence - obtain the general form of the intensity ratios of multiplets associated with antiphase single-quantum coherences, with an expression given in terms of the Jacobi polynomials.\par

An extension of the de Moivre-Laplace theorem, beyond the trivial case $q=0$, is applied to the $q$-th columns of the quantum Pascal pyramid, and is given in terms of a product of the $q$-th order Hermite polynomials and a Gaussian distribution, reproducing the well-known functional forms of the solutions of the quantum harmonic oscillator and the classical limit of Hermite-Gaussian modes in laser physics (Allen et al., \emph{Phys. Rev. A.}, \textbf{45}, 1992). This is used to approximate the Fourier-transformed spectra of $\hat{Z}_N^q$-associated multiplets of arbitrary complexity.\par

Finally, an exercise is shown in which the first two columns of the quantum Pascal pyramid are used to calculate the previously known symmetry-constrained upper bound on $I_z \rightarrow S_z$ polarization transfer in $I_N S$ spin systems.
\end{abstract}

\maketitle

\section{Introduction}
All magnetic resonance spectroscopists will be familiar with Pascal's triangle, which is invariably present in elementary courses on the interpretation of solution-state NMR or EPR spectra~\cite{mann_analysis_1995}. It is common knowledge that the $N$-th row of Pascal's triangle provides the intensities of a multiplet arising from J-coupling of the observed spin to a group of $N$ identical spin-1/2 nuclei.\par

Far less well-known is the "quantum Pascal pyramid" (a name suggested by OEIS \href{https://oeis.org/A268533}{A268533}) - a generalization of Pascal's triangle that arises in advanced multispin problems, providing a direct mapping between Cartesian product operators of a given spin product rank and single transition operators of a given quantum number, and providing the otherwise nonobvious intensity ratios of observable multispin single-quantum coherences.\par

To our knowledge, while structures of this type may be well-studied in the larger field of quantum dynamics, particularly at the intersection of quantum optics and laser physics~\cite{allen_orbital_1992,nienhuis_paraxial_1993,steely_harmonic_1997,enderlein_unified_2004,steuernagel_equivalence_2005,briggs_propagation_2024} (where they map Laguerre-Gaussian and Hermite-Gaussian modes), and even in the mathematics of superoscillations~\cite{colombo_evolution_2023} originally encountered by Aharonov in the weak value problem \cite{aharonov_how_1988,berry_evolution_2006,berry_roadmap_2019,aharonov_unified_2022}, they have not been previously considered in the magnetic resonance literature with the sole exception of Werbeck and Hansen's excellent relaxometry studies on the AX$_{4}$ spin system of $^{15}$NH$_{4}$~\cite{WERBECK2014136,hansen_measurement_2017}, who refer to a "modified Pascal's triangle". Also deserving of mention is the impressive recent paper by  Walder and Fritzsching~\cite{walder_weighted_2023} which described a closely-related generalization of the "$z^{k}$ multiplets", of which the $z^{0}$ and $z^{1}$ multiplets appear in this paper.

\section{Multispin operator algebra}
Consider a subspace with $N$ identical spin-$1/2$ particles. Within the product operator formalism commonly used in NMR~\cite{SORENSEN1984163}, Sørensen defined~\cite{sorensen_polarization_1989} a relevant operator $\hat{Z}_N^q$, defined as $2^q$ times the sum of the $\binom{N}{q}$ distinct longitudinal magnetization operators with spin product rank $q$. These terms arise from an expansion of the following cumulative tensor product:
\begin{equation}
\bigotimes_{i=1}^{N} \left(\hat{\mathbb{1}}+2\hat{I}_{iz}\right) = \sum_{q=0}^{N} \hat{Z}_{N}^q    
\end{equation}
And we have obtained an alternative expression for $\hat{Z}_N^q$:
\begin{equation}
\hat{Z}_{N}^q   = \frac{2^q}{q!} \mathlarger{\mathlarger{\sum}_{\{i_1...i_q\}=1}^{N}} (\epsilon_{i_1...i_N})^2 \bigotimes_{k=1}^{q} \hat{I}_{i_k z}
\end{equation}
In which $\epsilon_{i_{1}\dots i_{N}}$ refers to the $N$-dimensional Levi-Civita symbol.\par

Some explicit examples of $\hat{Z}_N^q$ are:
\begin{figure}[htbp!]
\begin{equation}
\begin{split}
&\hat{Z}_2^0 = \hat{\mathbb{1}}\\
&\hat{Z}_2^1 = 2\left(\hat{I}_{1z}+ \hat{I}_{2z}\right)\\
&\hat{Z}_2^2 = 4\left(\hat{I}_{1z}\hat{I}_{2z}\right)\\
\end{split}
\end{equation}
\caption{The set of the $\hat{Z}_{2}^q$ operators.}
\end{figure}

\begin{figure}[htbp!]
\centering
\begin{equation}
\begin{split}
&\hat{Z}_3^0 = \hat{\mathbb{1}}\\
&\hat{Z}_3^1 = 2\left(\hat{I}_{1z}+ \hat{I}_{2z}+\hat{I}_{3z}\right)\\
&\hat{Z}_3^2 = 4\left(\hat{I}_{1z}\hat{I}_{2z}+\hat{I}_{1z}\hat{I}_{3z}+\hat{I}_{2z}\hat{I}_{3z}\right)\\
&\hat{Z}_3^3 = 8\left(\hat{I}_{1z}\hat{I}_{2z}\hat{I}_{3z}\right)
\end{split}
\end{equation}
\caption{The set of the $\hat{Z}_{3}^q$ operators.}
\end{figure}

\begin{figure}[htbp!]
\centering
\begin{equation}
\begin{split}
\hat{Z}_4^0 &= \hat{\mathbb{1}}\\
\hat{Z}_4^1 &= 2\left(\hat{I}_{1z}+ \hat{I}_{2z}+\hat{I}_{3z}+\hat{I}_{4z}\right)\\
\hat{Z}_4^2 &= 4(\hat{I}_{1z}\hat{I}_{2z}+\hat{I}_{1z}\hat{I}_{3z}+\hat{I}_{1z}\hat{I}_{4z}\\&\qquad+\hat{I}_{2z}\hat{I}_{3z}+\hat{I}_{2z}\hat{I}_{4z}+\hat{I}_{3z}\hat{I}_{4z})\\
\hat{Z}_4^3 &= 8\left(\hat{I}_{1z}\hat{I}_{2z}\hat{I}_{3z}+\hat{I}_{1z}\hat{I}_{2z}\hat{I}_{4z}+\hat{I}_{1z}\hat{I}_{3z}\hat{I}_{4z}+\hat{I}_{2z}\hat{I}_{3z}\hat{I}_{4z}\right)\\
\hat{Z}_4^4 &= 16\left(\hat{I}_{1z}\hat{I}_{2z}\hat{I}_{3z}\hat{I}_{4z}\right)
\end{split}
\end{equation}
\caption{The set of $\hat{Z}_{4}^q$ operators.}
\end{figure}

And a relevant operator that will appear throughout this paper is the weighted variant of $\hat{Z}_N^q$:
\begin{equation}\label{eqn:reducedZcoefficient}
\hat{z}_N^q = \binom{N}{q}\hat{Z}_N^q
\end{equation}
\newpage
\section{The quantum Pascal pyramid and single-quantum coherences}
Consider the scenario in which $N$ identical spin-1/2 particles were now symmetrically \emph{coupled} to another "probe" spin species we will denote $S$. For the sake of explicitness, we will assume that the evolution is strictly under a secular J-coupling Hamiltonian:
\begin{equation}
H_J = 2\pi J_{IS} \hat{I}_{z}\hat{S}_z = \pi J_{IS} \hat{Z}_{N}^1 \hat{S}_{z}
\end{equation}

In general, all pure $\hat{Z}_N^{q} \hat{S}_{x}$ or $\hat{Z}_N^{q} \hat{S}_{y}$ operators - for example - correspond to single-quantum coherences and produce a multiplet that may be observed at the Larmor frequency of the $S$-spins. In more abstract terms, the detected multiplet intensities correspond to - and "read out" - the eigenvalue spectrum of $\hat{Z}_N^{q}$.

It is well-known that the muliplet intensities of the trivial case $\hat{Z}_N^{0}S_{x}\equiv S_{x}$ are given by the $N$th row of Pascal's triangle i.e. $\binom{N}{k}$, but the appearance of $\hat{Z}_N^{q}S_{x}$ multiplets is generally not as straightforward.

To resolve this predicament, it would be desirable to have operators explicitly describing the pure populations of states with magnetic quantum number $m$, which can be denoted $\hat{S}_{N}^m$. This is given by the elegant relations:
\begin{equation}\label{eqn:ZtoSmap}
\begin{split}
\hat{S}_N^m= \ \ \ \mathlarger{\sum}_ {q=0}^N \ \ \ &\hat{z}_N^q \left(\frac{\partial}{\partial x}\right)^{N/2-m} \frac{(1-x)^q (1+x)^{N-q}}{\left(N/2-m\right)!} \rvert_{x=0} \\
\hat{z}_N^q=\mathlarger{\sum}_ {m=-N/2}^{+N/2} &\hat{S}_N^m \left(\frac{\partial}{\partial x}\right)^{N/2-m} \frac{(1-x)^q (1+x)^{N-q}}{\left(N/2-m\right)!} \rvert_{x=0}
\\
\end{split}
\end{equation}
Exploiting the resemblance of Equation \ref{eqn:ZtoSmap} with the Rodrigues formula for the Jacobi polynomials $P_{n}^{\left(\alpha,\beta\right)}\left(x\right)$, we derived (rediscovering results known in quantum optics since Wunsche~\cite{wunsche_role_2003} and Abramochkin \& Volostnikov~\cite{abramochkin_generalized_2004}, if not earlier):
\begin{equation}\label{eqn:StoZmap}
\begin{split}
\hat{S}_N^m= \ \ \ \sum_{q=0}^N \ \ \ &\hat{z}_N^q P_{N/2-m}^{\left(N/2+m-q,-N-1\right)}(3)\\
\hat{z}_N^q= \sum_{m=-N/2}^{N/2} &\hat{S}_N^m P_{N/2-m}^{\left(N/2+m-q,-N-1\right)}(3)\\
\end{split}
\end{equation}
The map between the operators $\hat{S}_N^m$ and $\hat{Z}_N^k$ in Equations \ref{eqn:ZtoSmap} and \ref{eqn:StoZmap} produces a rich symmetrical pattern related to previous studies of the quantum mechanics of orbital angular momentum~\cite{allen_orbital_1992,nienhuis_paraxial_1993}. Some explicit examples are:
\newpage
\begin{figure}[htbp!]
\begin{equation}
\begin{blockarray}{ccc}
 & \hat{Z}_{1}^{0} & \hat{Z}_{1}^{1} \\
\begin{block}{c(cc)}
\hat{S}_{1}^{-\frac{1}{2}}\ \ & 1 & -1 \\
\hat{S}_{1}^{+\frac{1}{2}}\ \ & 1 & 1 \\
\end{block}
\end{blockarray}
\end{equation}
\caption{\nth{1} slice of the quantum Pascal pyramid ($N$ = 1, $I$ = 1/2).}
\end{figure}
\begin{figure}[htbp!]
\begin{equation}
\begin{blockarray}{cccc}
 & \hat{Z}_{2}^{0} & \hat{Z}_{2}^{1} & \hat{Z}_{2}^{2} \\
\begin{block}{c(ccc)}
\hat{S}_{2}^{-1}\ \ & 1 & -2 & 1 \\
\hat{S}_{2}^{0}\ \ \ \ & 2 & 0 & -2 \\
\hat{S}_{2}^{+1}\ \ & 1 & 2 & 1 \\
\end{block}
\end{blockarray}
\end{equation}
\caption{\nth{2} slice of the quantum Pascal pyramid ($N$ = 2, $I$ = 1/2).}
\end{figure}

\begin{figure}[htbp!]
    \centering
\begin{equation}
\begin{blockarray}{ccccc}
 & \hat{Z}_{3}^{0} & \hat{Z}_{3}^{1} & \hat{Z}_{3}^{2} & \hat{Z}_{3}^3 \\
\begin{block}{c(cccc)}
 \hat{S}_3^{-\frac{3}{2}}\ \ & 1 & -3 & 3 & -1 \\
 \hat{S}_3^{-\frac{1}{2}}\ \ & 3 & -3 & -3 & 3 \\
 \hat{S}_3^{+\frac{1}{2}}\ \ & 3 & 3 & -3 & -3 \\
 \hat{S}_3^{+\frac{3}{2}}\ \ & 1 & 3 & 3 & 1 \\
\end{block}
\end{blockarray}
\end{equation}
\caption{\nth{3} slice of the quantum Pascal pyramid ($N$ = 3, $I$ = 1/2).}
\end{figure}

\begin{figure}[htbp!]
\centering
\begin{equation}
\begin{blockarray}{cccccc}
 & \hat{Z}_4^0 & \hat{Z}_4^1 & \hat{Z}_4^2 & \hat{Z}_4^3 & \hat{Z}_4^4 \\
\begin{block}{c(ccccc)}
 \hat{S}_4^{-2}\ \ & 1 & -4 & 6 & -4 & 1 \\
 \hat{S}_4^{-1}\ \ & 4 & -8 & 0 & 8 & -4 \\
 \hat{S}_4^0\ \ \ \ & 6 & 0 & -12 & 0 & 6 \\
 \hat{S}_4^{+1}\ \ & 4 & 8 & 0 & -8 & -4 \\
 \hat{S}_4^{+2}\ \ & 1 & 4 & 6 & 4 & 1 \\
\end{block}
\end{blockarray}
\end{equation}
\caption{\nth{4} slice of the quantum Pascal pyramid ($N$ = 4, $I$ = 1/2).}
\end{figure}

\begin{figure}[htbp!]
\centering
\begin{equation}
\begin{blockarray}{cccccccc}
 & \hat{Z}_6^0 & \hat{Z}_6^1 & \hat{Z}_6^2 & \hat{Z}_6^3 & \hat{Z}_6^4 & \hat{Z}_6^5 & \hat{Z}_6^6 \\
\begin{block}{c(ccccccc)}
  \hat{S}_6^{-3}\ \ & 1 & -6 & 15 & -20 & 15 & -6 & 1 \\
 \hat{S}_6^{-2}\ \ & 6 & -24 & 30 & 0 & -30 & 24 & -6 \\
 \hat{S}_6^{-1}\ \ & 15 & -30 & -15 & 60 & -15 & -30 & 15 \\
 \hat{S}_6^0\ \ \ \ & 20 & 0 & -60 & 0 & 60 & 0 & -20 \\
 \hat{S}_6^{+1}\ \ & 15 & 30 & -15 & -60 & -15 & 30 & 15 \\
 \hat{S}_6^{+2}\ \ & 6 & 24 & 30 & 0 & -30 & -24 & -6 \\
 \hat{S}_6^{+3}\ \ & 1 & 6 & 15 & 20 & 15 & 6 & 1 \\
\end{block}
\end{blockarray}
\end{equation}
\caption{\nth{6} slice of the quantum Pascal pyramid ($N$ = 6, $I$ = 1/2).}
\end{figure}
Most readers will be familiar with the leftmost column ($\hat{Z}_N^0$) of the quantum Pascal pyramids and a few may recognize the column $\hat{Z}_N^{1}$ which describes the intensity ratios of the multiplets that occur following the famous INEPT experiment~\cite{morris_enhancement_1979,doddrell_enhancement_1981,PEGG198132}, although Walder and Fritzching~\cite{walder_weighted_2023} have pointed out that the discovery of $\hat{Z}_N^{1}$-associated multiplets historically predates the INEPT experiment~\cite{pachler_intensity_1977}.\par

The central columns of the Pascal pyramids, $\hat{Z}_N^{\lfloor{N/2}\rfloor}$ and $\hat{Z}_N^{\lceil{N/2}\rceil}$ (which are equivalent for $N \in 2\mathbb{Z}$), have been called the "Pauli Pascal triangles" in pedagogical introductions to quantum mechanics~\cite{horn_didactical_2006,horn_pauli_2007}.\par

We may define a "reduced" quantum Pascal pyramid with the coefficients of $\hat{z}_N^q$ rather than $\hat{Z}_{N}^q$ (see Equation \ref{eqn:reducedZcoefficient} and Figure \ref{fig:reducedqpp}), which has the possible advantage of simplifying the matrices by removing the $\binom{N}{q}$ factor. Intriguingly, the coefficients of $\hat{z}_N^{q}$ form an antisymmetric Catalan's triangle $\binom{N-1}{N/2-m}-\binom{N-1}{N/2-m-1}$ with a similar form to the seemingly unrelated triangle of Clebsch-Gordan multiplicities $\binom{N}{N/2-\ell}-\binom{N}{N/2-\ell-1}$~\cite{sabba_catalan_2024} providing the number of irreducible spin-$\ell$ representations in a group of $N$ spin-$1/2$ particles, the latter being known since at least the time of Wigner \cite{wigner_chapter_1959}. This mathematical coincidence is less disturbing when one considers the ubiquity of Catalan triangles in virtually any type of combinatorial problem~\cite{forder_problems_1961}.

\begin{figure}[htbp!]
\centering
\begin{equation}
\begin{blockarray}{ccccccc}
 & \hat{z}_5^0 & \hat{z}_5^1 & \hat{z}_5^2 & \hat{z}_5^3 & \hat{z}_5^4 & \hat{z}_5^5 \\
\begin{block}{c(cccccc)}
 \hat{P}_5^{-\frac{5}{2}}\ \ & 1 & -1 & 1 & -1 & 1 & -1 \\
 \hat{P}_5^{-\frac{3}{2}}\ \ & 5 & -3 & 1 & 1 & -3 & 5 \\
 \hat{P}_5^{-\frac{1}{2}}\ \ & 10 & -2 & -2 & 2 & 2 & -10 \\
 \hat{P}_5^{+\frac{1}{2}}\ \ & 10 & 2 & -2 & -2 & 2 & 10 \\
 \hat{P}_5^{+\frac{3}{2}}\ \ & 5 & 3 & 1 & -1 & -3 & -5 \\
 \hat{P}_5^{+\frac{5}{2}}\ \ & 1 & 1 & 1 & 1 & 1 & 1\\
\end{block}
\end{blockarray}
\end{equation}
\caption{\nth{5} slice of the "reduced" quantum Pascal pyramid for a 5-spin-1/2 system.}
\label{fig:reducedqpp}
\end{figure}

\begin{figure}
    \centering
    \includegraphics[width=0.73\linewidth]{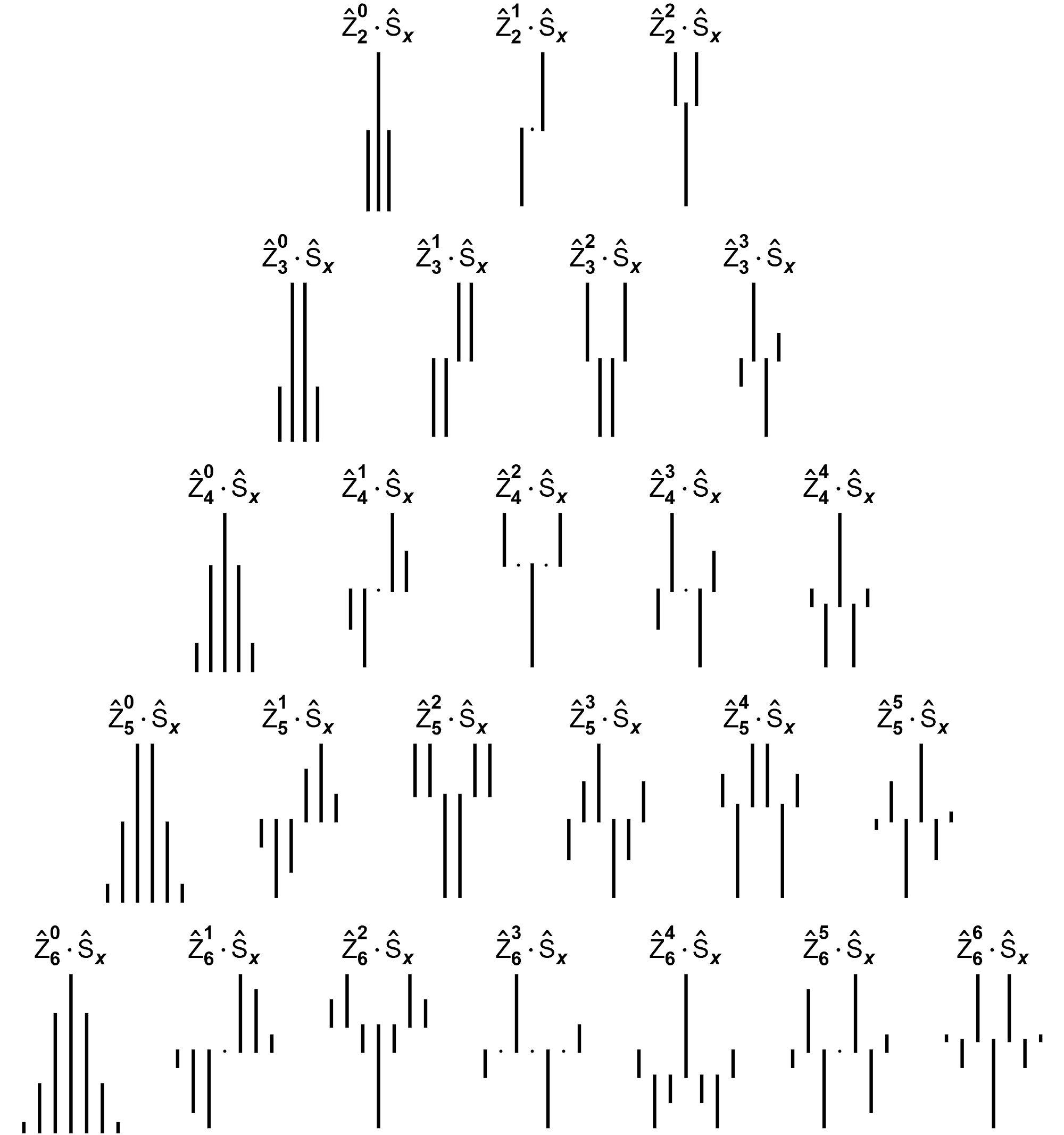}
    \caption{The quantum Pascal pyramid visualized as a triangle of the spectra of $\hat{Z}_N^q$ associated multiplets. The spectra have been scaled for equal visual visibility - the reader should \emph{be mindful} of the actual scaling factors $2^{-N} \binom{N}{q}$.}
\end{figure}

\section{Generalized quantum Pascal triangles}
The familiar Pascal's triangle (Table \ref{table:BinomialTriangle}), which we may call $\mathbb{T}^{(0)}$, can be generalized to a triangle $\mathbb{T}^{(q)}$ that describes $\hat{z}_N^{q}$-associated multiplets.

Begin by considering the properties of $\mathbb{T}^{(0)}$, which occurs in the series expansion of $(1+x)^N$:
\begin{equation}
\sum_{k=0}^{N} a_k^{(0)} x^k= (1+x)^N    
\end{equation}

The coefficients $a_k^{(0)}$ are given by the familiar binomial coefficients:
\begin{equation}
a_{k}^{(0)}=\binom{N}{k}
\end{equation}
\begin{table}[htbp!]
\begin{centering}
\begin{tabular}{c c c c c c c c c c c c c c c c c c c c c c c c c c c }
&$m_J$&\dots&-$\frac{5}{2}$&&-2&&-$\frac{3}{2}$&&-1&&-$\frac{1}{2}$&&0&&+$\frac{1}{2}$&&+1&&+$\frac{3}{2}$&&+2&&+$\frac{5}{2}$&&\dots\\
$N$\\
1 &&&&&&&&&&&1&&&&1&\\
2 &&&&&&&&&1&&&&2&&&&1&&&\\
3 &&&&&&&1&&&&3&&&&3&&&&1&&&\\
4 &&&&&1&&&&4&&&&6&&&&4&&&&1&\\
5 &&&1&&&&5&&&&10&&&&10&&&&5&&&&1\\
\vdots
\end{tabular}
\end{centering}
\caption{$\mathbb{T}^{0}$, the triangle of binomial coefficients, describing the multiplet intensities of the operator $\hat{z}_{N}^{0}\hat{S}_x$.}
\label{table:BinomialTriangle}
\end{table}
\\
The $\mathbb{T}^{(q)}$ triangles have a closely related generating function:
\begin{equation}
\sum_{k=0}^{N} a_k^{(q)} x^k= (1+x)^{N-q}(1-x)^{-q} 
\end{equation}
And an analytical expression for the coefficients of $a_k^{q}$ may be expressed in terms of Jacobi polynomials: 
\begin{equation}
a_k^{(q)}=P_{N-k}^{(k-q,-N-1)}(3)
\end{equation}
Some useful special cases are:
\begin{equation}
\begin{split}
a_k^{(1)}&= \binom{N-1}{k-1}-\binom{N-1}{k}\\
&=  \binom{N}{k}-2\binom{N-1}{k}
\end{split}
\end{equation}
\begin{equation}
a_k^{(2)} \ \ = \binom{N}{k}- 4 \binom{N-2}{k - 1}
\end{equation}
\begin{equation}
a_k^{(q)} = (-1)^{N+k}a_k^{(N-q)} 
\end{equation}

Several other potentially useful relations for $a_k^{(q)}$ may be found in the detailed mathematical analysis by Cação et al. \cite{cacao_combinatorial_2018}, where they are called "a family of Pascal trapezoids".

We will also define:
\begin{equation}
A_k^{(q)} = \binom{N}{q} a_{k}^{(q)}
\end{equation}
\begin{table}[htbp!]
\begin{centering}
\begin{tabular}{c c c c c c c c c c c c c c c c c c c c c c c c c c c }
&$m_J$&\dots&-$\frac{5}{2}$&&-2&&-$\frac{3}{2}$&&-1&&-$\frac{1}{2}$&&0&&+$\frac{1}{2}$&&+1&&+$\frac{3}{2}$&&+2&&+$\frac{5}{2}$&&\dots\\
$N$\\
1 &&&&&&&&&&&-1&&&&1&\\
2 &&&&&&&&&-1&&&&0&&&&1&&&\\
3 &&&&&&&-1&&&&-1&&&&1&&&&1&&&\\
4 &&&&&-1&&&&-2&&&&0&&&&2&&&&1&\\
5 &&&-1&&&&-3&&&&-2&&&&2&&&&3&&&&1\\
\vdots
\end{tabular}
\end{centering}
\caption{$\mathbb{T}^{1}$, the triangle of the coefficients $a_{k}^{(1)}$, describing the multiplet intensities of the operator $\hat{z}_{N}^{1}S_x$.}
\end{table}

\begin{table}[htbp!]
\begin{centering}
\begin{tabular}{c c c c c c c c c c c c c c c c c c c c c c c c c c c }
&$m_J$&\dots&-$\frac{5}{2}$&&-2&&-$\frac{3}{2}$&&-1&&-$\frac{1}{2}$&&0&&+$\frac{1}{2}$&&+1&&+$\frac{3}{2}$&&+2&&+$\frac{5}{2}$&&\dots\\
$N$\\
2 &&&&&&&&&1&&&&-2&&&&1&&&\\
3 &&&&&&&1&&&&-1&&&&-1&&&&1&&&\\
4 &&&&&1&&&&0&&&&-2&&&&0&&&&1&\\
5 &&&1&&&&1&&&&-2&&&&-2&&&&1&&&&1\\
\vdots
\end{tabular}
\end{centering}
\caption{$\mathbb{T}^{2}$, the triangle of the coefficients $a_{k}^{(2)}$, describing the multiplet intensities of the operator $\hat{z}_{N}^{2}S_x$.}
\end{table}

\begin{table}[htbp!]
\begin{centering}
\begin{tabular}{c c c c c c c c c c c c c c c c c c c c c c c c c c c c c c}
&$m_J$&\dots&-3&&-$\frac{5}{2}$&&-2&&-$\frac{3}{2}$&&-1&&-$\frac{1}{2}$&&0&&+$\frac{1}{2}$&&+1&&+$\frac{3}{2}$&&+2&&+$\frac{5}{2}$&&+3&&\dots\\
$N$\\
3 &&&&&&&&&-1&&&&3&&&&-3&&&&1&&&\\
4 &&&&&&&-1&&&&2&&&&0&&&&-2&&&&1&\\
5 &&&&&-1&&&&1&&&&2&&&&-2&&&&-1&&&&1\\
6 &&&-1&&&&0&&&&3&&&&0&&&&-3&&&&0&&&&1\\
\vdots
\end{tabular}
\end{centering}
\caption{$\mathbb{T}^{3}$, the triangle of the coefficients $a_{k}^{(3)}$, describing the multiplet intensities of the operator $\hat{z}_{N}^{3}S_x$.}
\end{table}
\newpage
\section{Generalization of the de Moivre-Laplace theorem}
\begin{figure}[tbhtbp!]
    \centering
    \includegraphics[width=0.75\linewidth]{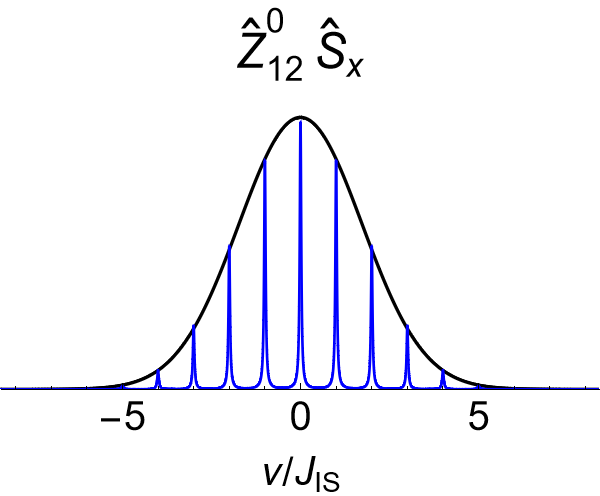}
    \caption{The spectrum $S_{12}^{(0)}$ (blue) 
 of the $\hat{Z}_{12}^{0}$-associated multiplet and its Gaussian approximation $G_{12}^{(0)}$ (black). A linewidth of $w=0.05 J_{IS}$ is assumed.}
    \label{fig:Z120 multiplets}
\end{figure}

The de Moivre-Laplace theorem is the famous, extraordinary mathematical result that a binomial distribution with $N$ elements converges to a Gaussian distribution as $N$ tends to infinity. 
\\
That is to say, a Gaussian distribution may be used to approximate a binomial distribution. In magnetic resonance terms, we could say that the familiar $\hat{Z}_N^0$-associated multiplet, for an $I_N S$ spin system, has a Fourier-transformed spectrum $S_N^{(0)}$ that is roughly traced by the Gaussian distribution $G_{N}^{(0)}$ (see Figure \ref{fig:Z120 multiplets}):
\begin{equation}\label{eqn:deMoivreLaplace0}
G_{N}^{(0)}\left(\nu\right) =  \left(\frac{2}{\pi w}\right)\sqrt{\frac{2}{\pi N}} \exp{\left(-\frac{2}{N} \left(\frac{\nu}{J_{IS}}\right)
   ^2\right)}
\end{equation}

In which $\nu$ is the spectral frequency and $w$ is the spectral linewidth; the $2/\left(\pi w\right)$ factor arises under the assumption that the spectrum $S_N^{(0)}$ consists of a normalized sum of $A_k^{q}$-weighted individual Lorentzians $L\left(\nu\right)$:
\begin{equation}
S_N^{(0)}\left(\nu\right) = 2^{-N}\mathlarger{\sum}_{m=-N/2}^{+N/2} A_{N/2+m}^{(0)} L\left(\nu\right)
\end{equation}
Where each Lorentzian is characterized by a centre frequency $\nu_0 = m J_{IS}$ at which the maximum intensity is $2/\left(\pi w\right)$:
\begin{equation}
L\left(\nu\right)=\left(\frac{1}{\pi}\right)\frac{\left(w/2\right)}{\left(\nu-m J_{IS}\right)^2+\left(w/2\right)^2}
\end{equation}

\begin{figure}[ht!]
    \centering
    \includegraphics[width=0.95\linewidth]{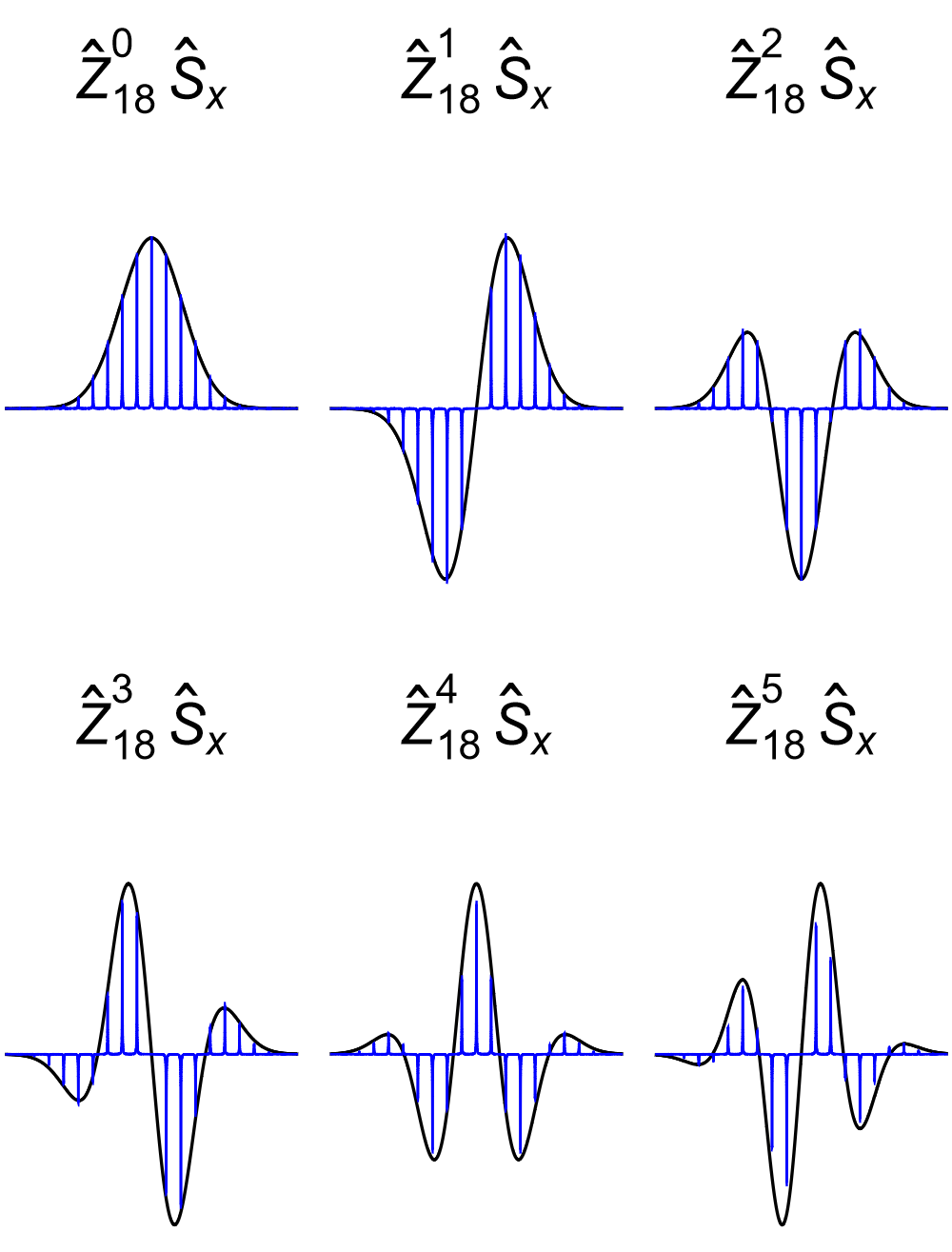}
    \caption{The spectra $S_{18}^{(q)}(\nu)$ (blue) of $\hat{Z}_{18}^{q}$-associated multiplets and the Hermite-Gaussian approximations $\hat{G}_{18}^{q}(\nu)$ (black). A line width of $w = 0.02 J_{IS}$ is assumed.}
    \label{fig:Z18q}
\end{figure}

Now, consider the general case of $\hat{Z}_N^{q}$-associated multiplets, whose spectra may be denoted $S_N^{(q)}\left(\nu\right)$:
\begin{equation}
S_N^{(q)}\left(\nu\right) = 2^{-N}\mathlarger{\sum}_{m=-N/2}^{+N/2} A_{N/2+m}^{(q)} L\left(\nu\right)
\end{equation}  
\\
Remarkably, the result of Equation \ref{eqn:deMoivreLaplace0} can be generalized to $Z_N^q$-associated multiplets, which closely correspond to distributions given by the $q$-th Gaussian derivatives $G_N^{(q)}\left(\nu\right)$:
\begin{equation}
G_{N}^{(q)}\left(\nu\right) = \frac{(-1)^q \left(N/2\right)^{q}}{q!} \frac{d ^q}{d \nu ^q} G_{N}^{(0)}\left(\nu\right)
\end{equation}
And, again exploiting the Rodrigues formula, $G_{N}^{(q)}$ may be expressed in terms of the well-known Hermite polynomials $H_q(x)$:
\begin{equation}\label{eqn:HermiteGauss}
G_{N}^{(q)}\left(\nu\right)=\frac{\left(N/2\right)^{q/2}}{q!} H_q\left(\sqrt{\frac{2}{N}}\nu\right) G_{N}^{(0)}\left(\nu\right)    
\end{equation}
The reader is encouraged to compare the form of Equation \ref{eqn:HermiteGauss} (see Figure \ref{fig:Z18q}) with the textbook case of the wavefunctions of the quantum harmonic oscillator, and the classical limit of Hermite-Gaussian laser modes~\cite{allen_orbital_1992,nienhuis_paraxial_1993,steely_harmonic_1997,steuernagel_equivalence_2005,enderlein_unified_2004,briggs_propagation_2024}. 

Derivatives are a measure of a rate of change, and - apart from the trivial relationship that $\hat{Z}_N^q$-associated multiplets change sign exactly $q$ times - there are at least two physical pictures associated with $\hat{Z}_N^q$ operators:
\begin{enumerate}
\item Each $\hat{Z}_N^{q}$ operator transforms with $\cos^q\left(\beta\right)$ under a rotation of flip angle $\beta$ in the $xy$-plane, defined through a rotation operator $\hat{R}_\varphi\left(\beta\right) = \exp\left[-i  \beta\left(\cos(\varphi)\hat{I}_x+\sin(\varphi)\hat{I}_y\right)\right]$:
\begin{equation}
\left\langle \hat{R}_\varphi\left(\beta\right)\hat{Z}_N^q \hat{R}_\varphi\left(-\beta\right) \rightarrow \hat{Z}_N^q \right\rangle=\frac{\left(\hat{R}_\varphi\left(\beta\right)\hat{Z}_N^q \hat{R}_\varphi\left(-\beta\right)\vert \hat{Z}_N^q\right)}{\left(\hat{Z}_N^q \vert \hat{Z}_N^q\right)} = \cos^{q}\left(\beta\right)
\end{equation}

Here, $\left(\hat{A}\vert\hat{B}\right)= \Tr\left[\hat{A}^\dag  \hat{B}\right]$ denotes the commonly used Liouville bracket, with curious readers referred to Pileio's excellent book \cite{pileio_lectures_2022}.

\item Each pure $\hat{Z}_N^{q}$ operator can be converted - with a $R_\varphi\left(\pi/2\right)$ rotation that may be realized by a simple $(90^{\circ})_0$ pulse - into $I$-spin multiple quantum coherence containing a mixture of coherence orders $p \in \left\{-q,-q+2\dots+q-2,+q\right\}$. If we use $\hat{\mathscr{Q}}_{q,p}$ to denote a $\hat{Z}_N^q$-derived operator of pure coherence order $p$, the following decomposition applies:
\begin{equation}
\hat{R}_\varphi\left(\pi/2\right)\hat{Z}_N^q \hat{R}_\varphi\left(-\pi/2\right) = 2^{-q}\mathlarger{\mathlarger{\sum}}_{\substack{p=-q \\ q+p \in 2\mathbb{Z}}}^{+q} \binom{q}{\frac{1}{2}(q-p)} \hat{\mathscr{Q}}_{q,p}    
\end{equation}

The \emph{maximum} (absolute) coherence order available is $\lvert\pm 
q\rvert$, and by definition~\cite{BODENHAUSEN1984370}, $q$-quantum coherences transform with $\exp\left[-i q\times \beta\right]$ under a $z$-rotation of flip angle $\beta$, defined through the rotation operator $\hat{R}_z\left(\beta\right) = \exp\left[-i  \beta \hat{I}_z\right]$:

\begin{equation}
\left\langle \hat{R}_z\left(\beta\right)\hat{\mathscr{Q}}_{q,p}  \hat{R}_z\left(-\beta\right) \rightarrow \hat{\mathscr{Q}}_{q,p}  \right\rangle = \exp\left[-i q\times \beta\right]
\end{equation}

Here, it should be noted that common NMR strategies for the excitation of $n$-quantum coherences in $I_n S$ systems often involve the generation of $Z_N^n$-associated spin order in the preparation and/or readout steps~\cite{huang_enhanced_2017}.

\end{enumerate}

\newpage
\section{Application: Bounds on Polarization Transfer}
 The problem of bounds on spin order transfer in magnetic resonance - a subset of the more general question of \emph{reachability} in unitarily controlled Markovian quantum systems \cite{ende_reachability_2020} - has been explored repeatedly~\cite{LEVITT19921,REDFIELD1991642,nielsen_2d_1992,stoustrup_generalized_1995,nielsen_accessible_1992,zhang_absence_1994,nielsen_conditional_1995,glaser_unitary_1998,untidt_design_1998,untidt_analytical_2000,levitt_symmetry_2016,pravdivtsev_symmetry_2022,mr-2-395-2021,BENGS2024107631} in the field of NMR since the seminal set of papers by Sørensen~\cite{sorensen_polarization_1989,sorensen_universal_1990,sorensen_entropy_1991}. At the more abstract level, this topic has also been shown to be intimately related to the mathematics of $C$-numerical ranges \cite{li_generalized_1984,li_cnumerical_1994,schulte-herbruggen_significance_2008,dirr_relative_2008}.\par
 
 Consider the following question: in an $I_NS$ spin system, what are the symmetry-constrained upper bounds on transfer of spin order from $\hat{I}_z$ to $\hat{S}_z$?\par

It is possible to answer this question and derive the bounds on $\hat{I}_z\rightarrow\hat{S}_z$ polarization transfer by using the quantum Pascal pyramid and the original Sørensen arguments~\cite{sorensen_polarization_1989}.\par

The first step of a polarization transfer experiment is typically the conversion of $I$-spin order to an intermediate of maximally $I$-correlated $S$-spin order, which is trivial to do in this particular case with a unitary transformation corresponding to the aforementioned INEPT pulse sequence~\cite{morris_enhancement_1979}: 
\begin{equation}
\hat{I}_z \ \ \xrightarrow{90_y^{I}} \ \ \xrightarrow[t=1/(2J_{IS})]{H_{J}}\ \ \xrightarrow{90_x^{I}}\ \ 2\hat{I}_z\hat{S}_z    
\end{equation}
Recalling that:
\begin{equation}
2\hat{I}_z\hat{S}_z =\hat{Z}_{N}^1\hat{S}_z
\end{equation}
And having obtained the expressions for the (eigenvalue) spectrum $\lambda(\hat{A})$ of the relevant operators:
\begin{equation}
\lambda\left(\hat{Z}_N^0\right) \in \left\{A_{0}^{(0)}\dots A_{N}^{(0)}\right\}=\left\{1,N\dots N,1\right\}
\end{equation}
\begin{equation}
\lambda\left(\hat{Z}_N^1\right) \in \times\left\{A_{0}^{(1)}\dots A_{N}^{(1)}\right\} = \left\{-N,N(2-N)\dots N(N-2),N\right\}
\end{equation}
We can calculate the trace (i.e. the sum of eigenvalues) of these operators:
\begin{equation}
\Tr\left(\hat{Z}_N^{0}\right)=\sum_{k=0}^{N} A_k^{(0)} = 2^N
\end{equation}
\begin{equation}
\Tr\left(\hat{Z}_N^{1}\right)=\sum_{k=0}^{N} A_k^{(1)} = 0
\end{equation}

As is obvious from its antisymmetric form, $\hat{Z}_N^1$ is traceless. In magnetic resonance terms, one would say that $I$-spin decoupling (which collapses the multiplet) leads to no detectable $\hat{Z}_N^1 S_x$ signal, since the individual multiplet peak amplitudes, given by $\lambda\left(\hat{Z}_N^1\right)$, obviously add up to zero.\par

Nevertheless, we may appreciate that while unitary transformations cannot create or destroy spin order, they can certainly rearrange it. Suppose that one subjected $\hat{Z}_N^{1}$ to an ideal refocusing transformation (which can take, for example, the form of selective pulses~\cite{roy_enhancement_2015} or pulsed analogues thereof~\cite{NIELSEN1989359,sorensen_polarization_1989}) such that it was converted to an analogous operator $\left\vert\hat{Z}_N^1 \right\vert$ with strictly positive eigenvalues:
\begin{equation}
\lambda\left(\left\vert\hat{Z}_N^1 \right\vert\right) \in \left\{\left\vert A_{0}^{(1)}\right\vert\dots \left\vert A_{N}^{(1)}\right\vert\right\}=\left\{N,N(N-2)\dots N(N-2),N\right\}
\end{equation}
The trace of this operator has a simple closed-form expression:
\begin{equation}
\Tr\left(\left\vert\hat{Z}_N^1 \right\vert\right)=\sum_{k=0}^{N} \left\vert A_k^{(1)} \right\vert = 2 N\binom{N-1}{\left\lfloor N/2\right\rfloor} 
\end{equation}
\\
The magnitude of the total detected signal, $s(\hat{A}.\hat{S}_x)$, for a given $I$-spin longitudinal operator $\hat{A}$, is proportional to the gyromagnetic ratio of the "source" nucleus (from which spin order originates), and the trace of $\hat{A}$. \\
By evaluating the nominal, "unenhanced" signal, of $S_x$:
\begin{equation}
s\left(\hat{Z}_N^{0}\hat{S}_x\right) \propto \left\vert\gamma_{S}\right\vert \Tr\left(\hat{Z}_N^{0}\right)
\end{equation}
And the maximally enhanced signal of refocused $I$-correlated spin order:
\begin{equation}
s\left(\left\vert\hat{Z}_N^1 \right\vert\hat{S}_x\right) \propto \left\vert\gamma_{I}\right\vert  \Tr\left(\left\vert\hat{Z}_N^1 \right\vert\right)
\end{equation}

We may obtain the maximum possible enhancement factor in terms of $k_\gamma=\left\vert \gamma_I/\gamma_S\right\vert $:
\begin{equation}
\begin{split}
b_{\rm{max}}\left(\hat{I}_z\rightarrow \hat{S}_z\right) &= \frac{s\left(\left\vert\hat{Z}_N^1 \right\vert\hat{S}_x\right)}{s\left(\hat{Z}_N^{0}\hat{S}_x\right)} \\&= k_\gamma\frac{\Tr\left(\left\vert\hat{Z}_N^1 \right\vert\right)}{\Tr\left(\hat{Z}_N^0\right)}\\&=k_\gamma 2^{1-N} N \binom{N-1}{\lfloor{N/2}\rfloor}
\end{split}
\end{equation}
Which is a form of Sørensen's original expression~\cite{sorensen_polarization_1989}, and can be expressed in other elegant forms that have not previously appeared in the NMR literature:
\begin{equation}
\begin{split}
&= k_\gamma\times \begin{cases}
 \frac{N\text{!!}}{(N-1)\text{!!}} & \rm{for}~N\in 2 \mathbb{Z}+1 \\
 \frac{(N-1)\text{!!}}{(N-2)\text{!!}} & \rm{for}~N\in 2 \mathbb{Z}
\end{cases}
\\
&= 2 k_\gamma\times  MD_N
\end{split}
\end{equation}
Here, $!!$ denotes the double factorial while $MD_N$ is the mean absolute deviation of a symmetric binomial distribution with $N$ elements, as first discovered by de Moivre in the 1730 work \emph{Miscellanea Analytica}, a story discussed in detail by Diaconis and Zabell~\cite{diaconis_closed_1991}.\par

In a strange accident of history, it was de Moivre's fateful encounter with the latter $MD_N$ problem that led him to an early form of the central limit theorem (first appearing in \emph{The Doctrine of Chances} in 1738) a result so far ahead of its time that it was only rediscovered almost a century later by Laplace, the basis of the modern nomenclature "de Moivre-Laplace theorem". Towards the end of the last century, generalizations of this result were exploited by mathematicians \cite{diaconis_closed_1991} to prove various relationships involving - among other aspects - the Jacobi and Hermite polynomials that appear in this paper. It may be concluded that in the search for deeper patterns within quantum angular momentum, we must have done little more than retraced de Moivre's old steps, perhaps with an element of time reversal.

\section*{acknowledgements}
I am grateful to Malcolm H. Levitt for his generous support and mentorship over the years, for introducing me to the fascinating problem of bounds on spin order transfer, and for encouraging the creative environment in his group that led to this work. The author thanks Christian Bengs, Harry Harbor-Collins, and James W. Whipham for their interest in (and patience with) some ideas within this article. While this research was conducted in a personal capacity, the author acknowledges support from the European Research Council (Grant No. 786707-FunMagResBeacons).

\printbibliography

\end{document}